\documentclass{webofc}
\usepackage[varg]{txfonts}   
\woctitle{QUARKS-2016 19th International Seminar on High Energy Physics}
\begin{document}
\title{Charmed particles production in pA-interactions at 70 GeV}
%
%

\author{\firstname{Vasily} \lastname{Ryadovikov}\inst{1}\fnsep\thanks{\email{riadovikov@ihep.ru}} \\
(on behalf of the SVD Collaboration)
}
\institute{Institute for High Energy Physics, Sq. Nauki 1, Protvino, Moscow region 142281, Russia}

\abstract{The results of the SERP-E-184 experiment at the U-70 accelerator (IHEP, Protvino)
are presented. Interactions of the 70 GeV proton beam with carbon, silicon and lead targets
were studied to detect decays of charmed $D^0$, $\bar D^0$, $D^+$, $D^-$ mesons and $\Lambda_c^+$ 
baryon near their
production threshold. Measurements of lifetimes and masses have shown a good agreement with
PDG data. The inclusive cross sections of charm production and their A-dependencies have been
obtained. The yields of these particles are compared with the theoretical predictions and
the data of other experiments. The measured cross section of the total open charm production
$\sigma(c\bar c) = 7.1 \pm 2.3(stat) \pm 1.4(syst)$ $\mu$b/nucleon at the collision c.m. 
energy $\surd s$ = 11.8 GeV
is well above the QCD model predictions. The contributions of different kinds of charmed 
particles to the total cross section of the open charm production in proton-nucleus interactions
vary with energy.
}
\maketitle

\section{Introduction}
We represent the results of data processing of E-184 experiment on studying of charmed
particles production mechanisms in pA-interactions at 70 GeV. The experiment is carried out
by Cooperation SINP MSU (Moscow) – JINR (Dubna) – IHEP (Protvino) at the U-70 accelerator
in Protvino. Importance of work is that:\\
- The majority of experiments on a charm are executed with the electron beams where
their main properties are studied (measurements of mass, decay branching etc.).\\
- The experiments with hadron and heavy ion beams gave an opportunity to study the mechanisms
of charmed particle production in varying nuclear media from threshold energies
to energies of LHC.\\
- The charm production by hadrons is sensitive to the medium modifications. The charmed particles
are good probes to investigate properties of this medium, which may show up as quark-gluon plasma.\\
- More then 20 years ago in IHEP the measurements of cross sections of the charmed particle
production at the near-threshold energy region were performed by beam-dump experiment with
an absorber of muons \cite{Astratyan}, by the SCAT bubble chamber experiment \cite{Ammosov} and by the experiment
with BIS-2 spectrometer \cite{Aleev1}. The measured total cross sections in the energy range of
the primary beam from 40 to 70 GeV have proved to be much higher than the model predictions
based on QCD.\\
Experiment of E-184 is carried out on the Spectrometer with Vertex Detector setup (SVD).

\section{SVD setup detectors}
The main elements of the setup for charm searching \cite{Avdeichikov} are the high-precision
micro-strip vertex detector (MSVD) with an active target (AT) and a magnetic
spectrometer (MS) of the aperture equal to 1.8 x 1.2 m2 and the field of 1.18 T
within the region 3 m long. The active target (AT) (Fig. 1) contains 5 Si-detectors
each 300 $\mu$m thick and 1-mm pitch strips, the Pb-plate (220 $\mu$m thick) and C-plate
(500 $\mu$m thick). The micro-strip tracking part of MSVD consists of 10 Si-detectors.
\begin{figure*}[h]
\centering
\includegraphics[width=6cm,clip]{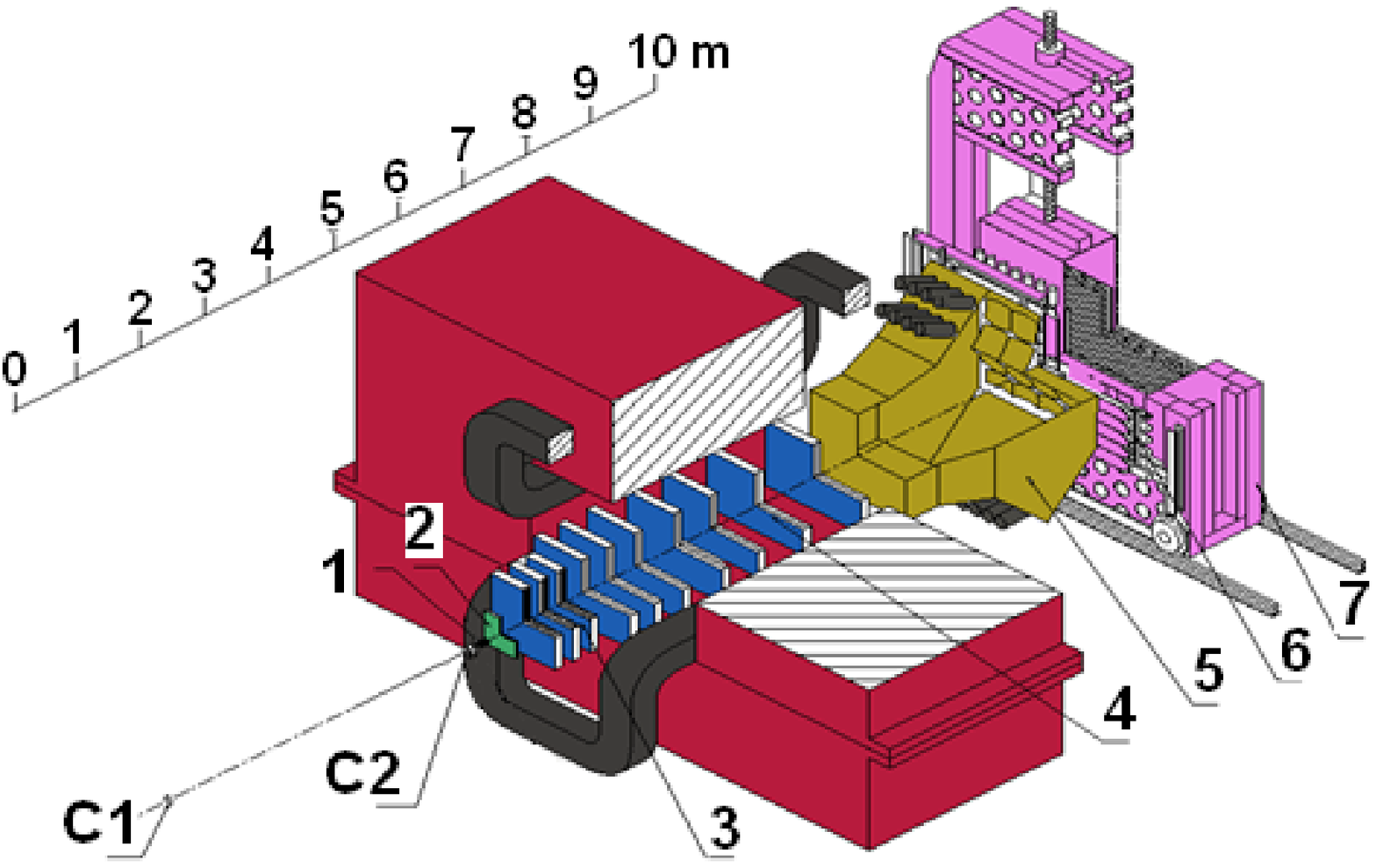}
\includegraphics[width=7cm,clip]{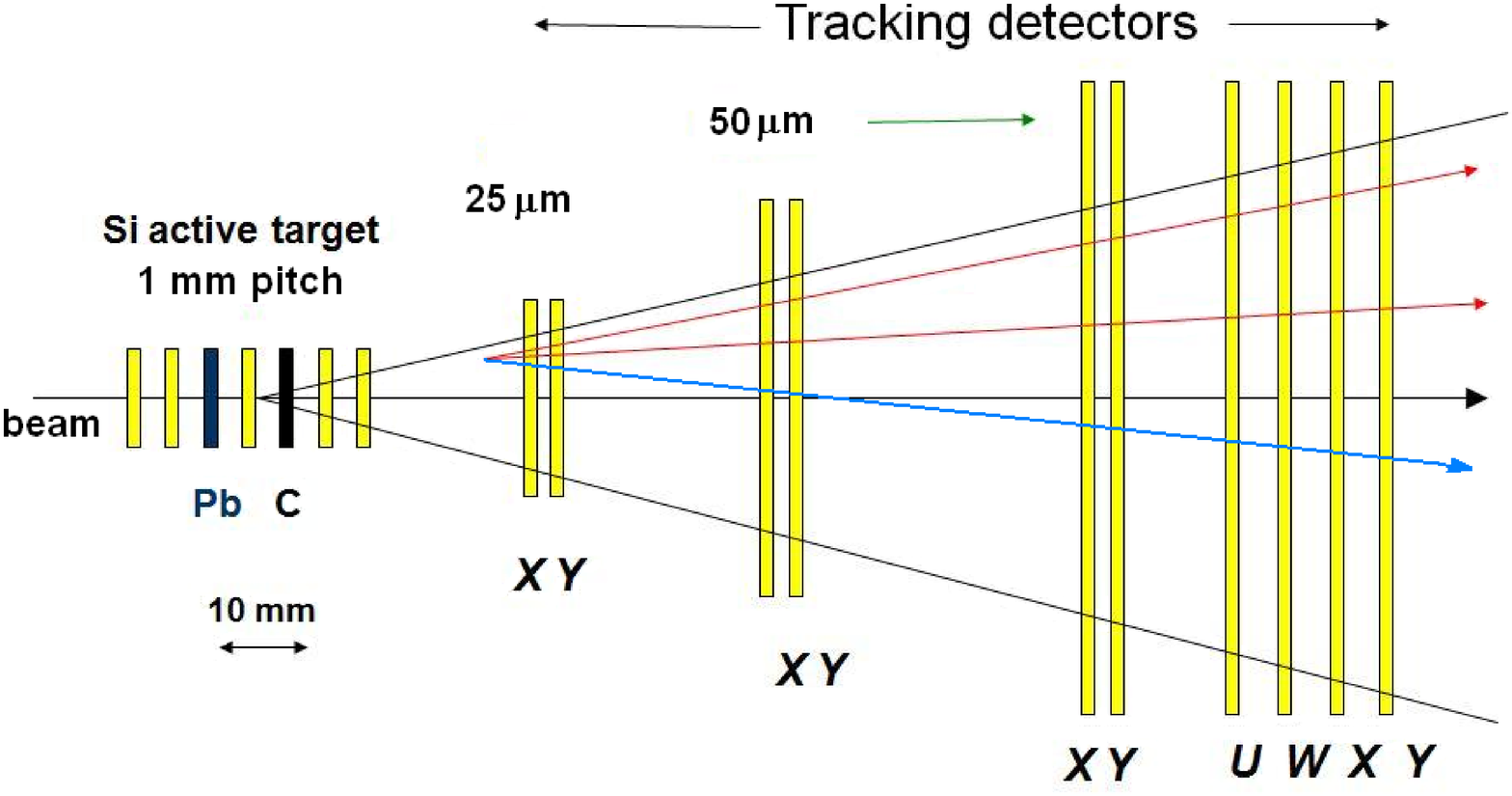}
\caption{SVD-2 layout and scheme of  MSVD. 
C1,C2 - beam scintillation counters; 1 - Si active target (AT); 
2 - microstrip vertex Si-detector (MSVD); 3, 4 - MWPC of  magnetic spectrometer (MS); 
5 - threshold Cherenkov counter (CC); 6 - scintillation hodoscope (SC);
7 - detector of gamma-quanta (DEGA).}
\label{fig-1}       
\end{figure*}

\begin{figure}[h]
\centering
\includegraphics[width=7cm,clip]{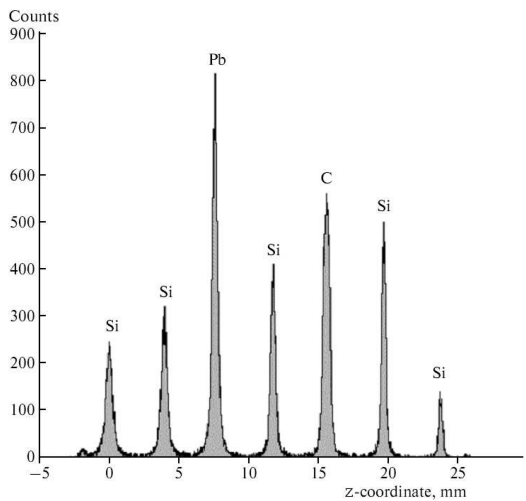}
\caption{The reconstructed Z-coordinates of the primary vertices in AT.}
\label{fig-2}       
\end{figure}

The vertex detector plays the main role in search of secondary vertexes with possible
decays of a charm as the distance between a point of interaction and D meson decay
point doesn't exceed 2 mm. We have the accuracy of primary vertex reconstruction along
a beam of 100 microns (Fig. 2), secondary vertex – 200 microns. The coordinate accuracy is
equal to 10 microns in the cross plane. The spectrometer features allow one to get 
the effective mass resolution of $\sigma$ = 4.4 MeV/$c^2$ for $K_s^0$ and 1.6 MeV/$c^2$ 
for $\Lambda^0$ masses,
for the particles decayed before the MSVD tracking detectors. 

\section{Simulations and data processing}

Modeling is needed for:\\
- to optimize of event selection criteria. As a result of modeling selection
criteria of events with the minimum background were taken;\\
- to define the detection efficiency for a charm particle;\\
- to confidence that we have really charm particles by the comparison of parameters
(decay length, etc.) of MC and experimental data of charm particles decays.\\

Proton-nucleus interactions have been simulated by means of FRITIOF7.02 code \cite{FRITIOF}.
The Fermi motion of nucleons, the deformation of nucleus and multiple re-scattering
are taken into account. The nucleon distribution density in a nucleus is described
in our case by the Woods–Saxon potential 
$\rho (r) = \rho (0)/(1 + exp[(r - r(0) \cdot A^{1/3}]/c)$,
with
$r(0) = 1.16(1 - 1.16A^{-2/3})$ fm and c = 0.5 fm. Hadronization processes described by the Lund
scheme using the fragmentation function $f(z) \sim z^{-1}(1 - z)^\alpha \cdot exp(-bm^2_t/z)$. 
The parameters
of this function were chosen as a = 0.18 and b = 0.34 $GeV^{-2}$ in accordance with the results
of the $e^+ e^-$ experiments OPAL \cite{Alexander} and \cite{Huang} where the parameters were adjusted by
the measured spectra of $D$ and $D^*$ mesons. Default values were used for the remaining
parameters in the FRITIOF code.

GEANT3.21 package \cite{GEANT} was used to simulate registration of pA-interactions.
The geometry of active and passive elements of the setup have been defined by means
of metrological measurements and corrected with results of the “straight tracks”
software alignment. The measured grid map of the magnetic field was applied. Charge
spreading over the MSVD strips, noises and cutoff amplitudes were introduced
channel-by-channel in accordance with experimentally measured values and data acquisition
parameters. The actual efficiencies obtained experimentally for the proportional wire
chambers were used for the magnetic spectrometer.

The simulated events have been processed by really using data handling system of
the SERP-E-184 experiment. The common data processing procedure started with the
filtration of MSVD data and reconstruction of tracks and primary vertices \cite{Kiriakov}.
Next, we selected events with the secondary vertices close to the interaction points
as the candidates for charm-production events. For this purpose the method of an 
analysis in the space of track parameters \cite{Vorobiev} was used. In this space each track
is presented with a point and all points for the tracks from the same vertex are 
located on a straight line. The results of simulation were compared with the 
experimental data. A good agreement was found for the numbers of minimum bias 
events in each of AT plates, for multiplicities of charged particles in primary 
vertices and for their momenta.

In case of $D^0 (\bar D^0) \rightarrow K \pi$ process the detection of $K^0_s \rightarrow \pi \pi$ decay in the MSVD can serve
as a reference procedure, because the well-known kaon production cross section is
many times larger than for charmed particles. It has been used to estimate the
detection efficiency of $V^0$ decay near the primary vertex and validate the data
processing algorithms for $D^0 (\bar D^0)$ \cite{Ryadovikov1}.

Additionally, the background minimum bias events were simulated without charm production.
This procedure was necessary to estimate the background conditions. The characteristics
of three-prong systems ($K \pi \pi$ and $pK \pi$) for MC events were compared to the experimental
data \cite{Ryadovikov2}, \cite{Ryadovikov3}. There is a good agreement between the simulated and experimental distributions
for a path length, momentum and x$_F$ variable (x$_F = p_L / \surd s$ in c.m.s.).

\section{Selection of events with charmed particles}
We have 52 million inelastic events in AT (C, Si, Pb).

\subsection{$D^0 \rightarrow K^- \pi ^+$, $\bar D^0 \rightarrow K^+ \pi ^-$ decays}
The candidates for the events with $D^0$ or $\bar D^0$ particle and its decay into $K \pi$ system
were selected using the following criteria \cite{Ryadovikov1}:

(i) The distance between the primary vertex and the $V^0$ vertex is more than 0.5 mm.

(ii) The decay tracks of the $V^0$ particle have non-zero impact parameters with
respect to the primary vertex, and the track of $V^0$ particle points to the primary vertex.

(iii) The effective mass of the $K\pi$ system differs from the world-average value
of the $D^0$ mass (1.865 GeV/c$^2$) by less than 0.5 GeV/c$^2$.

(iv) The momentum of the $K\pi$ system is higher than 10 GeV/c.

(v) The transverse momentum of the decay particle with respect to the direction
of motion of the $K\pi$ system is higher than 0.3 GeV/c. It follows from the analysis
of the Armenteros–Podolansky criterion and suppresses the background from neutral
kaons and $\Lambda ^0$ hyperons decays (Fig. 3 left).

(vi) Having a rather small number of events, the candidates for the ($D^0$/$\bar D^0$) 
particles were visually inspected by means of a specially designed graphic package on an high
resolution PC screen. In this way, hits which were not included into the reconstructed tracks in MSVD 
are tested. Using those hits and the visual picture of the event helps to eliminate obviously the background 
events with beam tracks, secondary interactions at AT, pile-up and ghost tracks.

\begin{figure*}[h]
\centering
\includegraphics[width=6cm,clip]{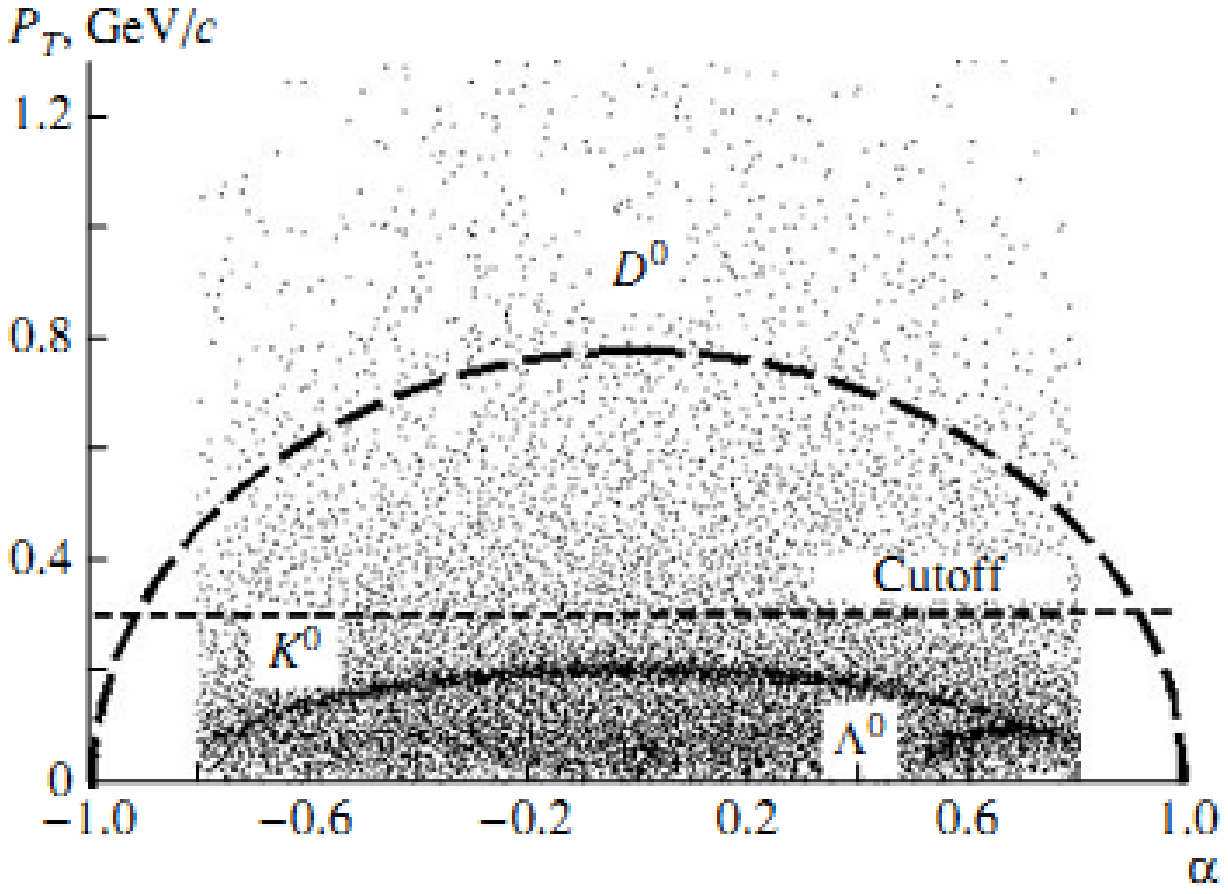}
\includegraphics[width=7cm,clip]{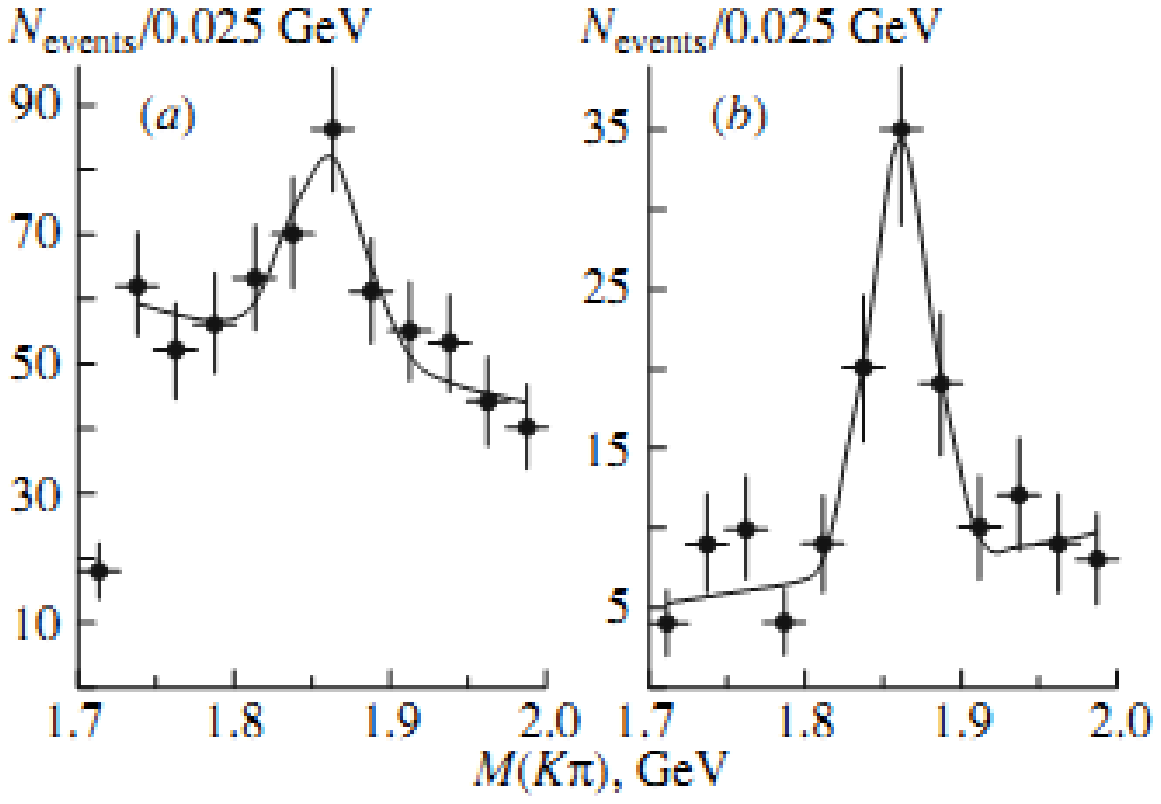}
\caption{ Left: Armenteros–Podolansky plot for 
$K^0, \Lambda ^0$ and $D^0$; $\alpha$ = ($P_L^+ - P_L^-$)/($P_L^+ + P_L^-$). 
Right: Effective mass spectra of the $K \pi$ system (a) before and (b) after visual inspection. }
\label{fig-3}       
\end{figure*}

The effective mass spectra of the $K\pi$ system after applying the criteria (i-v) and visual
inspection of the events are presented in Figs. 3a and 3b (right). The mass interval for the visual
inspection has been limited to the region from 1.7 to 2.0 GeV/c$^2$. The visual inspection
significantly reduces the background in the region of interest, but the charm + anti-charm
($D^0$/$\bar D^0$) signal decreased only on 20\%.	 The data fit by the sum of the straight
line and the Gaussian function is shown in Fig. 3b (right). It gives 1861 MeV/c$^2$ for
($D^0$/$\bar D^0$) mass with a standard deviation $\sigma$ = 21 MeV/c$^2$ ($\chi ^2$/NDF = 5.5/6) 
and the signal
to noise ratio of (51 $\pm$ 17)/(38 $\pm$ 13). The detection efficiency of ($D^0$/$\bar D^0$) particles
with efficiency of visual inspection taken into account is equal to
$\varepsilon (D^0 / \bar D^0)$ = 0.036.

\subsection{$D^+ \rightarrow K^- \pi ^+ \pi ^+$, $D^- \rightarrow K^+ \pi ^- \pi ^-$ decays}
For charged mesons we analyzed the ($K \pi \pi$) systems: 
$D^+ \rightarrow K^- \pi ^+ \pi ^+$, $D^- \rightarrow K^+ \pi ^- \pi ^-$.
The charged charmed mesons were found by analyzing the events with a three-prong secondary
vertex. The selection procedure of events went as follows (more details in \cite{Ryadovikov2}):

(i) Search for the third track to be associated with a two-prong secondary vertex
taking into account charges and kinematical correspondence to interaction vertex.

(ii) Selection of the $K \pi \pi$ systems with momentum P>7 GeV/c.

(iii) Cutting out non-physical regions and those with the highest background
using Dalitz plot for M($K \pi _1$) and M($K \pi _2$) variables.

(iv)  Within three tracks systems, two possible hypotheses for $K^0$ have shown a significant
peak in the mass distribution. Most of background was cut out after applying the condition
M$(\pi ^+ \pi ^-)_1$ + M$(\pi ^+ \pi ^-)_2$ < 1.2MeV/c$^2$.

(v) Hypotheses for ($K^- K^+ \pi$) system (possible $D_S$) were discarded with the 
M($K^- K^+ \pi$) > 1.93 MeV/c$^2$ cut.

(vi) Simulations have shown concentration of background events at the smallest decay
lengths, the condition L > 0.12 mm was applied to eliminate it (L = L$_{vis}$*M/P).

The mass distributions for $D^+$ and $D^-$ that was obtained with primary selection (i, ii) 
are shown in Fig. 4. Thay have significant background contributions, eliminated with
the rest of cuts (see Fig. 5).

\begin{figure}[h]
\centering
\includegraphics[width=13cm,clip]{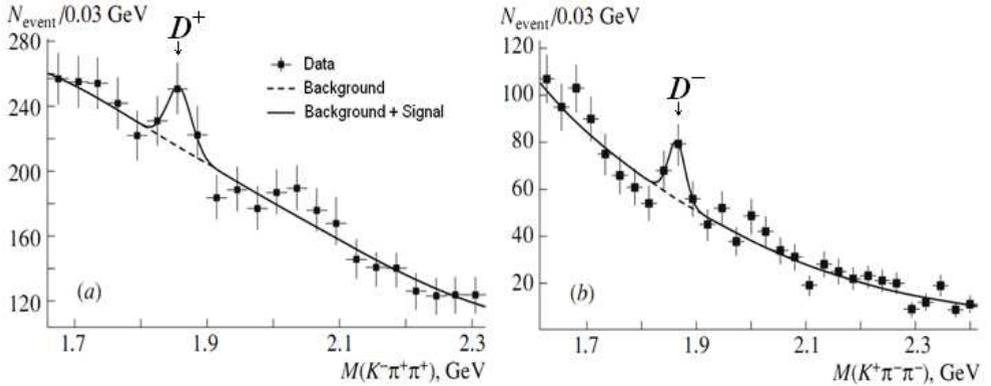}
\caption{ Effective mass spectra of the $K^- \pi ^+ \pi ^+$ (a) and $K^+ \pi ^- \pi ^-$ 
(b) systems after applying the conditions (i, ii). Data fitted with the sum of a Gaussian
function and a sixth order polynomial.  }
\label{fig-4}       
\end{figure}

\begin{figure*}[h]
\centering
\includegraphics[width=7cm,clip]{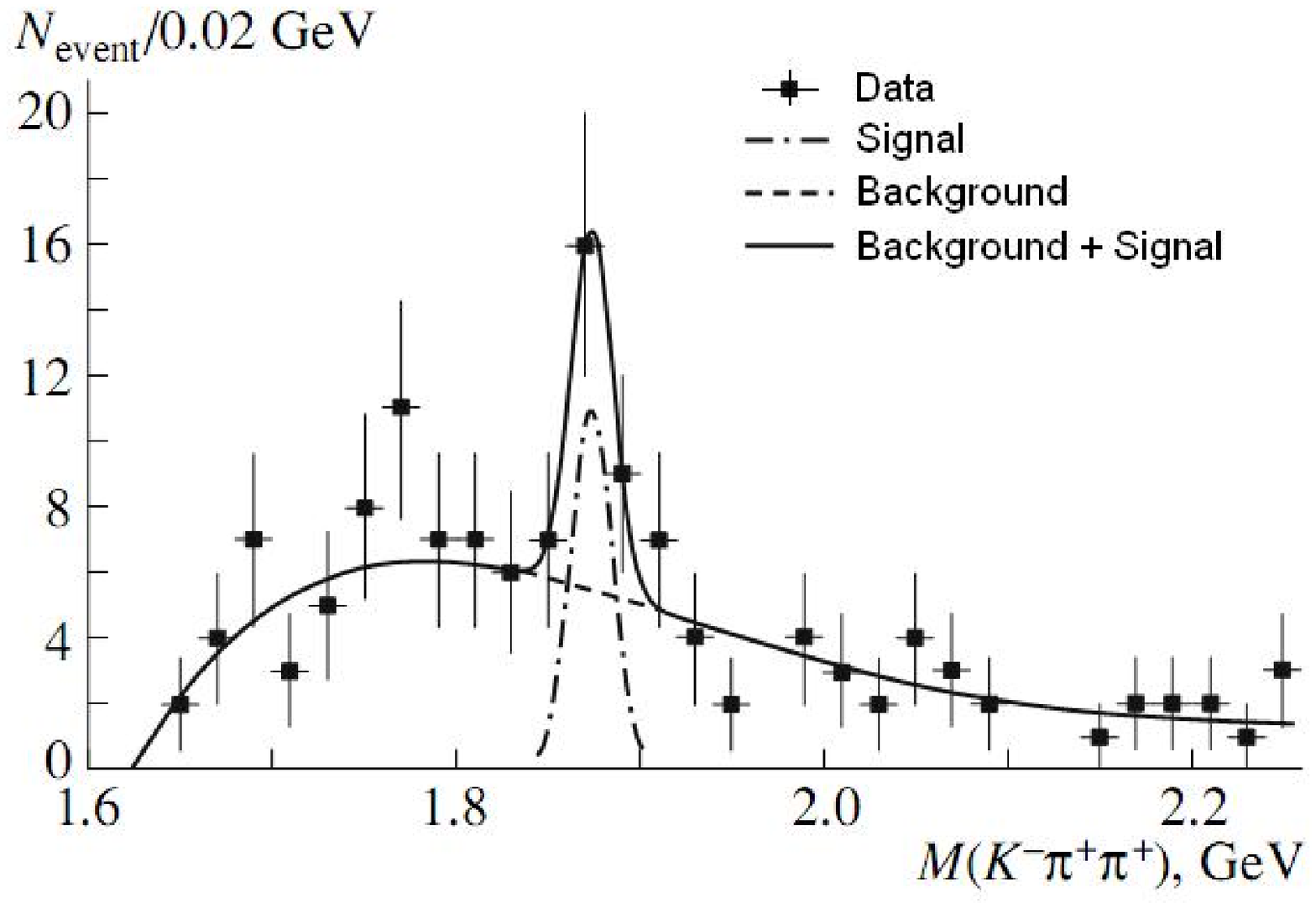}
\includegraphics[width=7cm,clip]{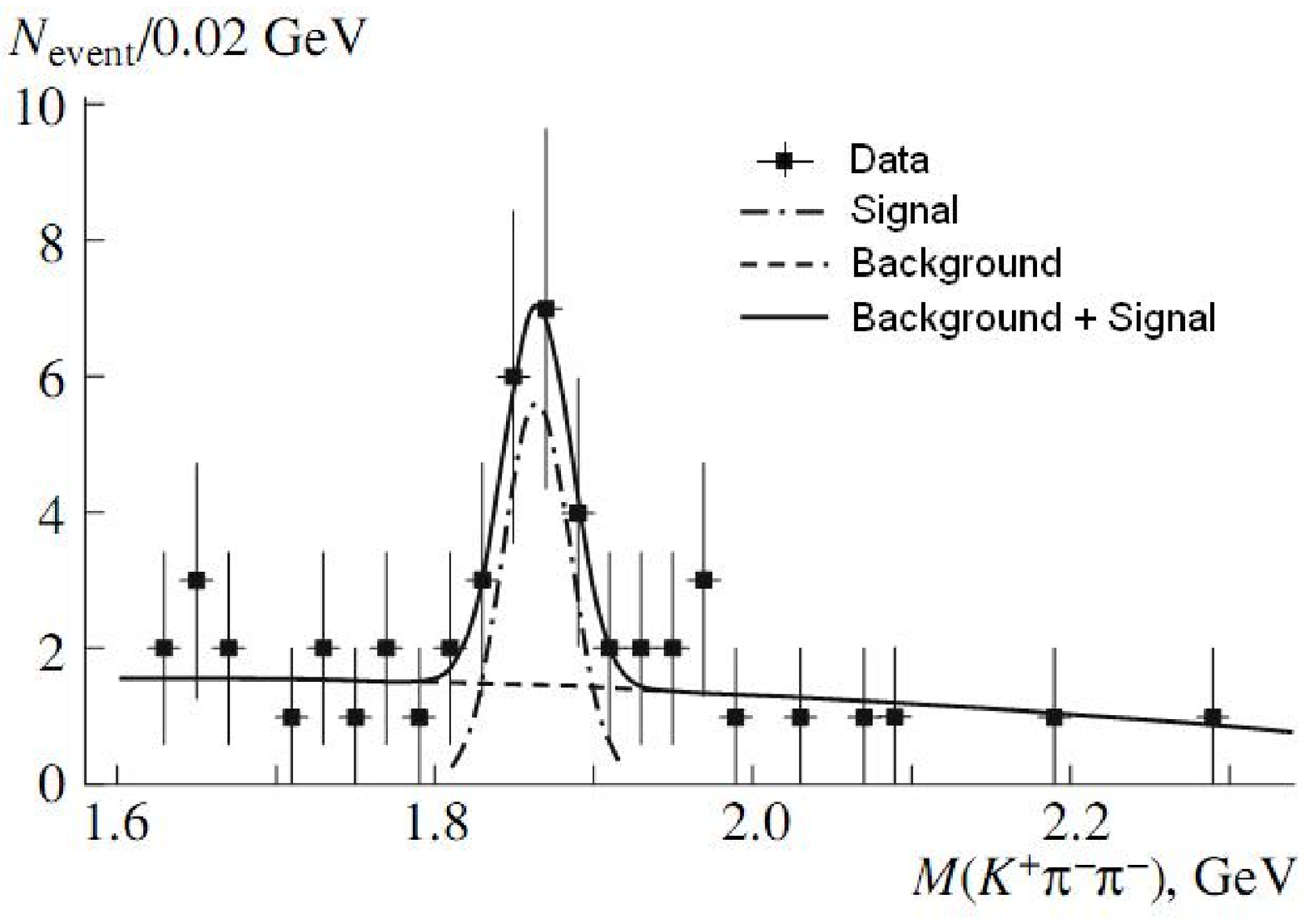}
\caption{ Effective mass spectra of the $K^- \pi ^+ \pi ^+$ (left) and $K^+ \pi ^- \pi ^-$ 
(right) systems after using all selection criteria.  }
\label{fig-5}       
\end{figure*}

After parameterizing the spectrum in Fig. 5 (left) in terms of the sum of the Gaussian function
and a fifth order polynomial ($\chi ^2$/NDF = 13.5/30), we got 15.5 $\pm$ 5.6 signal events from
the $D^+$ meson decay over the background of 16.6 $\pm$ 6.0 events. The measured $D^+$-meson
mass is 1874 $\pm$ 5 MeV/c$^2$ (the world-average value is 1869.6 MeV/c$^2$) with a standard
deviation of 11.5 MeV/c$^2$. The detection efficiency $\varepsilon (D^+)$ = 0.014 was defined from
simulations. The same procedure has been applied to search for the $D^-$ meson signal
(Fig. 5 right) in the mass spectrum of the $K^+ \pi ^- \pi ^-$. 
The spectrum was
parameterized as a sum of the Gaussian function and the second order
polynomial ($\chi ^2$/NDF = 3.6/20). The number of events in the signal was 15.0 $\pm$ 4.7 over
the background of 8.7 $\pm$ 2.7 events. The $D^-$-meson mass was 1864 $\pm$ 8 MeV/c$^2$, with
the standard deviation of 22 MeV/c$^2$. The detection efficiency obtained through
the simulation for the $D^-$ meson signal was equal to $\varepsilon (D^-)$ = 0.008.

\subsection{$\Lambda ^+ _c \rightarrow p K^- \pi ^+$  decays}

The charmed $\Lambda ^+ _c$ baryon was analyzed with the three-prong decay 
$\Lambda ^+ _c \rightarrow p K^- \pi ^+$.
The primary procedures were similar to those for the $D ^ \pm$ mesons decays (i, ii).
Without of particle identification, we had two hypotheses for positively
charged particles of the ($p K^- \pi ^+$) system in the effective mass spectrum (Fig. 6 left).
The next selection procedure of the events was carried out as follows (more details 
in \cite{Ryadovikov3}):

(iii) Dalitz plot for MC events with M($K^- \pi ^+$) and M($K^- p$) variables were used
in order to cut out the experimental data background.

(iv) To eliminate the $K^0$ background, it was required to have for the three-particle 
system M(3$\pi$) > 1.2 GeV/c$^2$. According to simulations, it cuts out about 85\% of
the background and only 12\% of the signal.

(v) The $p K^- \pi ^+$ system momentum met the following requirement: 25 < P < 60 GeV/c.

\begin{figure*}[h]
\centering
\includegraphics[width=7cm,clip]{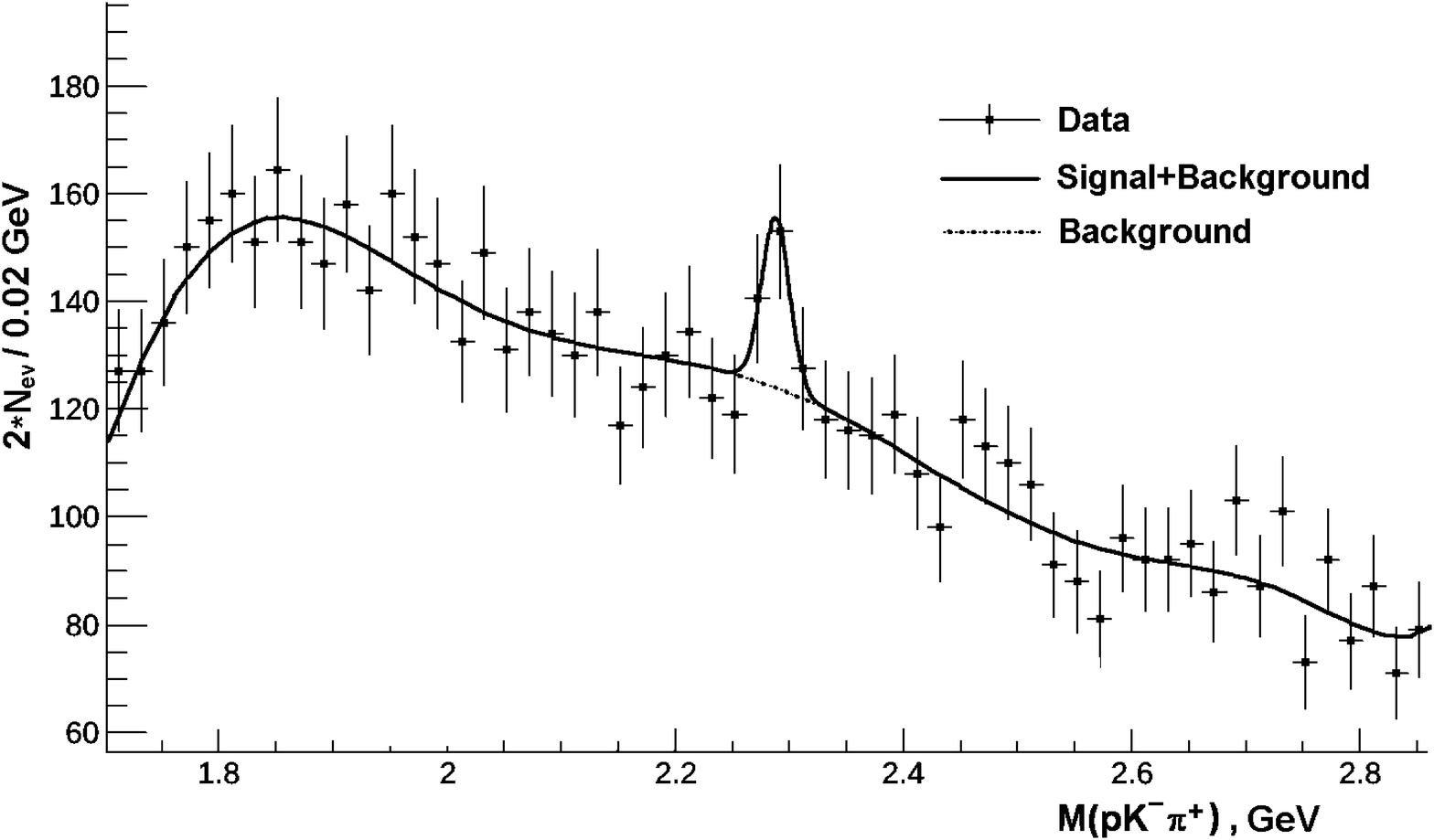}
\includegraphics[width=7cm,clip]{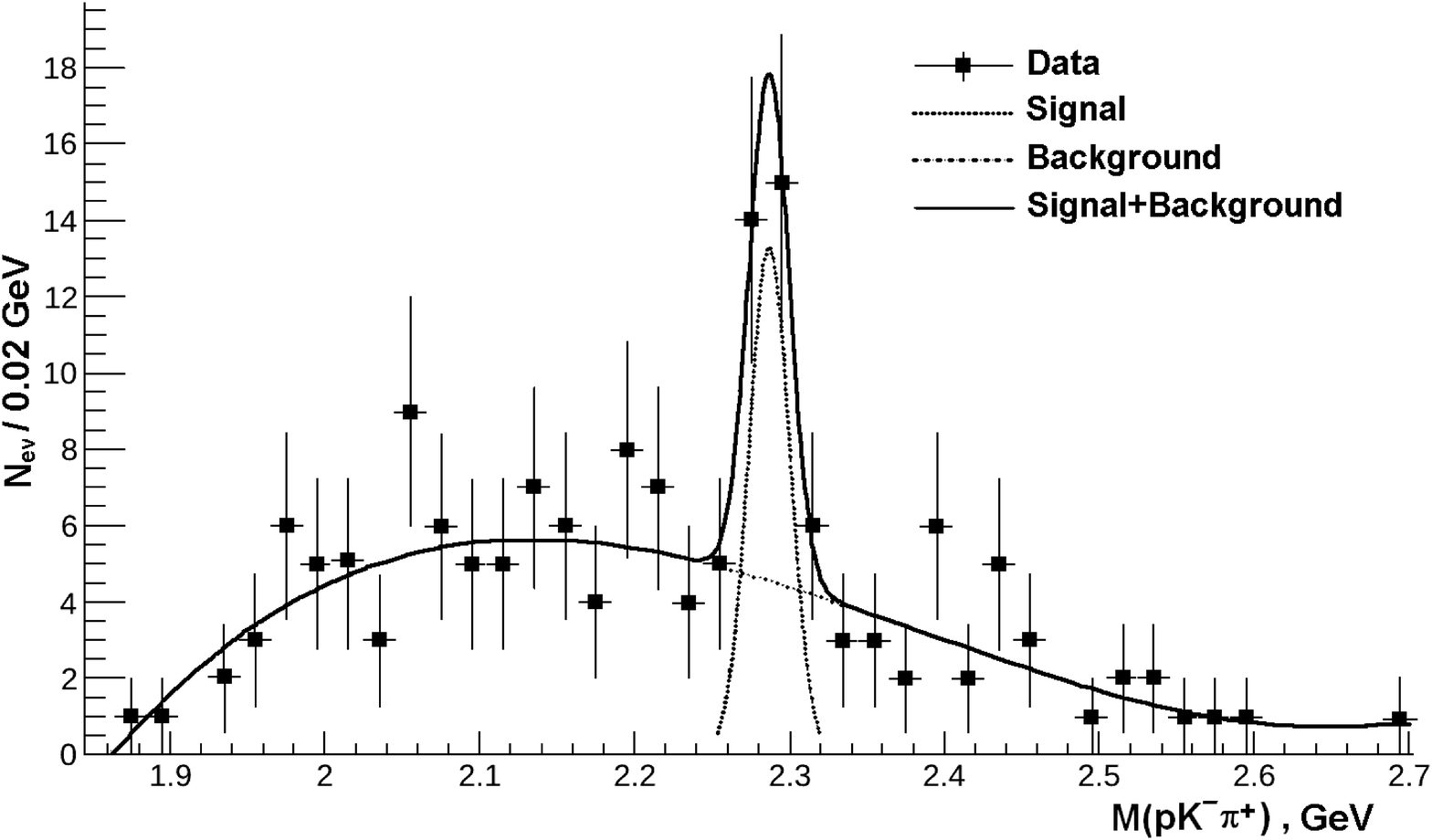}
\caption{ Effective mass spectra of the $p K^- \pi ^+$ system (two hypotheses) after 
the primary selection of events (left) and 
after using all selection criteria (right).  }
\label{fig-6}       
\end{figure*}

The application of all the criteria resulted in the effective mass spectrum is shown
in Fig. 6 right, with $\Lambda ^+ _c$ signal of 21.6 $\pm$ 6.0 events over the background of 16 $\pm$ 4
events,
the mass 2287 $\pm$ 4 MeV/c$^2$ and $\sigma$ = 13.1 MeV/c$^2$. In case when both of hypotheses turned
out to be in the peak region, they are taken with weights of 0.5 (there were 5\% of
such events in simulations and no one in the experimental data). The effective mass
spectrum of the $p K^- \pi ^+$ system has been parameterized by the sum of the Gaussian function
and the polynomial of the third degree ($\chi ^2$/NDF = 12.7/33). The detection
efficiency for $\Lambda ^+ _c \rightarrow p K^- \pi ^+$ was equal $\varepsilon$ = 0.011. 
The summary results of fitting the effective
mass charmed particles spectra are presented in Table 1.

\begin{table}
\centering
\caption{The results of effective mass spectra fitting.}
\label{tab-1}       
\begin{tabular}{|l|c|c|c|c|c|}
\hline
Particle decay & Signal & Background & $\chi ^2$/NDF 
& Detection & Mass, GeV/c$^2$   \\
& events & events & 
& efficiency & ($PDG mass$)  \\\hline
$D^0 \rightarrow K \pi$ & 51$\pm$17 & 38$\pm$13 & 5.5/6 & 0.036 & 1861$\pm$7 (\it {1864.8})  \\\hline
$D^+ \rightarrow K^-\pi^+\pi^+$ & 15.5$\pm$5.6 & 16.6$\pm$6.0 
& 13.5/30 & 0.014 & 1874$\pm$5 (\it {1869.6})  \\\hline
$D^- \rightarrow K^+\pi^-\pi^-$ & 15.0$\pm$4.7 & 8.7$\pm$2.7 & 3.6/20 & 0.008 
& 1864$\pm$8 (\it {1869.6})  \\\hline
$\Lambda ^+ _c \rightarrow pK^-\pi^+$ & 21.6$\pm$6.0 & 16$\pm$4 & 12.7/33 & 0.011 
& 2287$\pm$4 (\it {2286.5})  \\\hline
\end{tabular}
\end{table}

\section{Cross sections for charmed particle production and their A-dependence}
The following relation has been used to calculate the inclusive cross sections
for a given charmed particle:
\begin{center}
$N_s(i) = [N_0 \times (\sigma (i) \times A^ \alpha)/(\sigma_{pp} \times A^{0.7})] \times 
[(B(i) \times  \varepsilon (i))/K_{tr}]$, 
\end{center}
where $i = D^0, \bar D^0, D^+, D^-$ or $\Lambda ^+ _c$; $N_s(i)$ is the number of events
in the signal for ($i$) type of the charmed particle produced in the given target;
$N_0$ is the number of inelastic interactions in this target; $\sigma (i)$ is the cross section
for charmed particle ($i$) production at a single nucleon of the target; A is the atomic
mass number of AT material (C, Si or Pb); $\alpha (i)$ is an exponent parameter in
A-dependence of the charm cross section; $\sigma _{pp}$ is the cross section of the inelastic
proton-proton interaction at 70 GeV (= 31440 $\mu$b); $B(i)$ is the branching ratio for
the charmed particle decay 
($B(D^0/\bar D^0) \rightarrow K \pi$) = 0.038, $B(D^ \pm \rightarrow K \pi \pi$) = 0.094, 
$B(\Lambda ^+ _c \rightarrow p K^- \pi ^+)$ = 0.05); $\varepsilon (i)$ is the detection
efficiency of the charmed particle from Table 1 and $K_{tr}$ = 0.57 is the trigger efficiency
of registration of  inelastic events in our experiment.

Substituting $C(i) = [N_0/(\sigma_{pp} \times A^{0.7})]\times[B(i) \times \varepsilon (i)/K_{tr}]$, 
the relation takes the following form: 
$N_s(i) = C(i) \times \sigma (i) \times A^ \alpha$ or 
ln$(N_s(i)) = \alpha \times $ln$(A) + $ln$(\sigma (i))$. Figure 7 shows A-dependences of the charmed
particles production in AT.

\begin{figure}[h]
\centering
\includegraphics[width=7cm,clip]{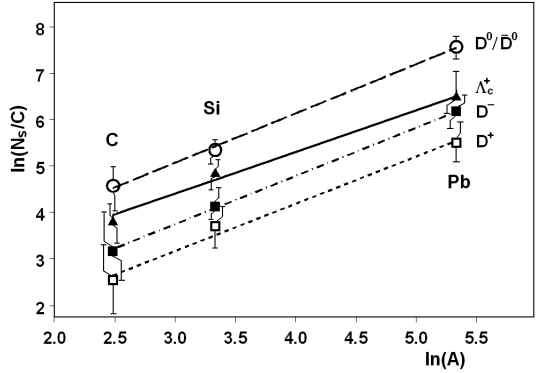}
\caption{ The A-dependence of cross sections for the charmed particles production in
pA-interactions. }
\label{fig-7}       
\end{figure}

The parameter $\alpha$ obtained from the straight linear approximations in Fig. 7 for each
particle is presented in Table 2 together with inclusive cross section and path length.
The $\alpha$-parameters are close to 1 for all charmed particles, as it was found earlier
for the hidden charm ($J/\psi$ and $\psi '$) production cross sections in the proton-nucleus
interactions \cite{Badier}, \cite{Alde}, \cite{Abreu}, \cite{Leitch}. 

\begin{table}
\centering
\caption{Characteristics of the charmed particles production. The first error is statistical, 
the  second – the systematical one.}
\label{tab-2}       
\begin{tabular}{|c|c|c|c|c|c|}
\hline
Num & Type of & Inclusive cross section & &\multicolumn{2}{c|} {c$\tau$, mm}   \\
 & charmed & for all x$_F$ & $\alpha$-parameter &\multicolumn{2}{c|} { }   \\ \cline{5-6}
 & particle & ($\mu$b/nucleon & & SVD-2 & PDG  \\\hline
 1 & $D^+$ & 1.2$\pm$0.4$\pm$0.2 & 1.02$\pm$0.26 & 0.291$\pm$0.075 & 0.311  \\\hline
 2 & $D^-$ & 1.9$\pm$0.6$\pm$0.4 & 1.04$\pm$0.27 & 0.341$\pm$0.088 & 0.311  \\\hline
 3 & $D^0$ & 2.5$\pm$0.8$\pm$0.5 & & &   \\ \cline{1-3}
 4 & $\bar D^0$ & 4.6$\pm$1.6$\pm$0.9 & 1.08$\pm$0.12 & 0.123$\pm$0.024 & 0.124   \\ \cline{1-3}
 5 & $D^0$/$\bar D^0$ & 7.1$\pm$1.8$\pm$0.9 & & &   \\\hline \cline{1-3}
 6 & $\Lambda ^+ _c$ & 4.0$\pm$1.6$\pm$1.2 & 0.9$\pm$0.2 & 0.051$\pm$0.011 & 0.059  \\\hline
\end{tabular}
\end{table}

On the basis of results presented in Table 2 and using the relation \cite{Andronic1}                     
\begin{center}
$\sigma _{tot}(c \bar c)$ = 0.5$\cdot$($\sigma_{D+} + \sigma_{D0} + \sigma_{D-} + \sigma_{\bar D0}
 + \sigma_{\Lambda c+} + \sigma_{Ds} + \sigma_{\bar Ds}$) 
\end{center}
total cross section of the charmed particles production in proton-nucleon interactions
at 70 GeV can be estimated as $\sigma(c \bar c) = 7.1 \pm 2.3(stat)\pm 1.4(syst)$
$\mu$b/nucleon. The charmed
particles yields measured in our experiment are given in Table 3 and in Fig. 8
along with the data from other experiments and theoretical predictions. 
                                                                                                                                                                                                                                                                                                                                                                                                              
\begin{table}
\centering
\caption{The charmed particles yields, $\sigma(i)/\sigma_{tot}(c \bar c)$.}
\label{tab-3}       
\begin{tabular}{|c|c|c|c|c|c|}
\hline
Yields $\backslash$ & PYTHIA & FRITIOF & SVD-2 &\multicolumn{2}{c|} {Other experiments}   \\
\cline{5-6}
particle & pp & pA & pA & NA-27 \cite{Aguilar} & HERA-B \cite{Abt}  \\\hline
$D^0$ & 0.28 & 0.51 & 0.35$\pm$0.16 & 0.57$\pm$0.08 & 0.44$\pm$0.18  \\\hline
$\bar D^0$ & 0.74 & 0.59 & 0.65$\pm$0.31 & 0.43$\pm$0.09 & 0.54$\pm$0.23  \\\hline
$D^+$ & 0.13 & 0.29 & 0.16$\pm$0.07 & 0.31$\pm$0.06 & 0.19$\pm$0.08  \\\hline
$D^-$ & 0.24 & 0.27 & 0.27$\pm$0.17 & 0.34$\pm$0.06 & 0.25$\pm$0.11  \\\hline
$\Lambda ^+ _c$ & 0.55 & 0.36 & 0.56$\pm$0.27 &
\multicolumn{2}{l|} {0.52$\pm$0.35 |  0.42$\pm$0.13 |  0.18$\pm$0.01}   \\
 & & & &\multicolumn{2}{c|} 
 {BIS-2 \cite{Aleev3} | E-769 \cite{Engelfried} | SELEX-2 \cite{Garcia}}   \\\hline
\end{tabular}
\end{table}

\begin{figure}[h]
\centering
\includegraphics[width=7cm,clip]{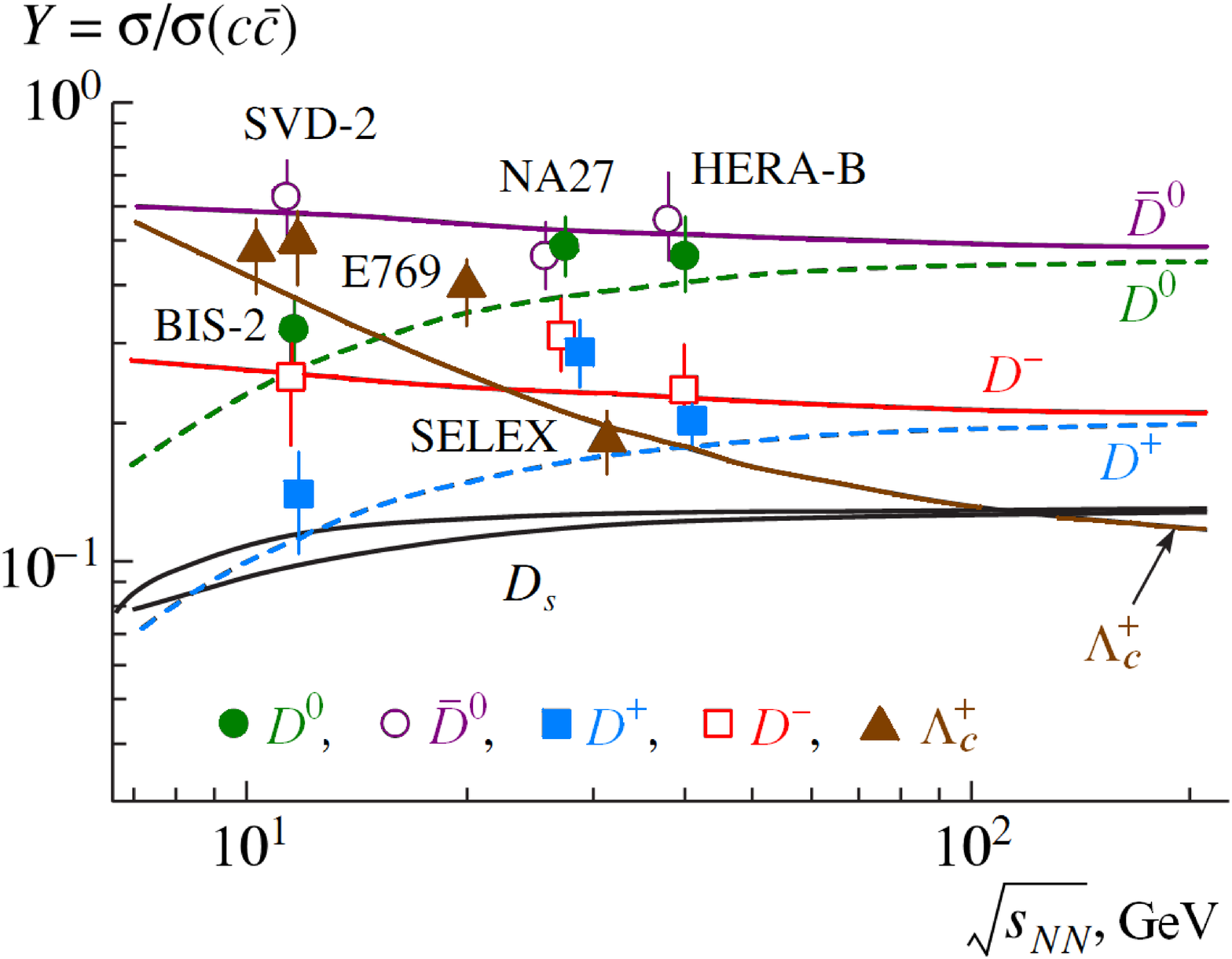}
\caption{ Relative yields of charmed particles. The experimental points
are taken
from Table 3, the theoretical curves (with designation of a particle) are taken 
from \cite{Andronic1}. }
\label{fig-8}       
\end{figure}

The contributions of charmed particles to the total cross section vary with energy.
For example, the particles ($D^0$ and $D^+$) contributions go down as the interaction
energy decreases to 70 GeV, while the antiparticles
($\bar D^0, D^-$ and $\Lambda ^+ _c$) contributions grow. A large difference of the charmed
particle and antiparticle yields was firstly observed experimentally in neutron–nucleus
interactions at the average neutron-beam energy of 43 GeV in the BIS-2 experiment \cite{Aleev4}.                                                                                                                
Only antiparticle ($\bar D^0$ and $D^-$) decays were observed there, while particle
($D^0$ and $D^+$) decays were not found (their cross sections proved to be below
the sensitivity threshold). The results shown in Fig. 8 are compatible with the predictions
of the statistical hadronization model \cite{Andronic1}, \cite{Andronic2}. 

\section{Results}
We present total cross section for open charm production and cross sections for the
inclusive production of $D$ mesons and $\Lambda ^+ _c$ baryon.
\begin{itemize}
\item The total open charm production cross-section at the collision c.m. energy
$\surd s$ = 11.8 GeV is  well above QCD models predictions and inclusive cross sections
are close to the prediction of QGSM for $D$ mesons and for $\Lambda ^+ _c$ baryon at
this energy (Fig. 9 left).
\begin{figure*}[hb]
\centering
\includegraphics[width=5.5cm,clip]{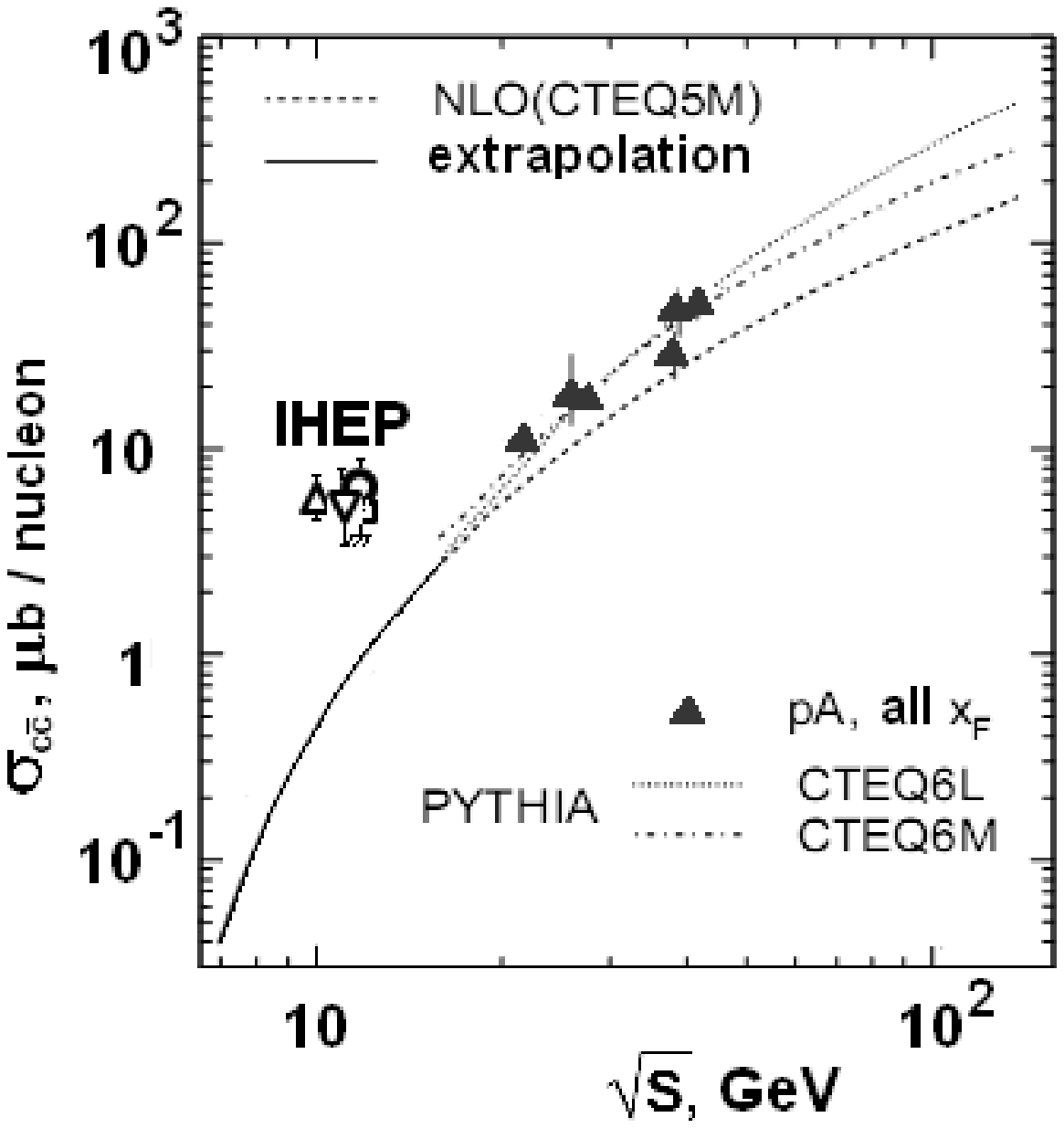}
\includegraphics[width=5.cm,clip]{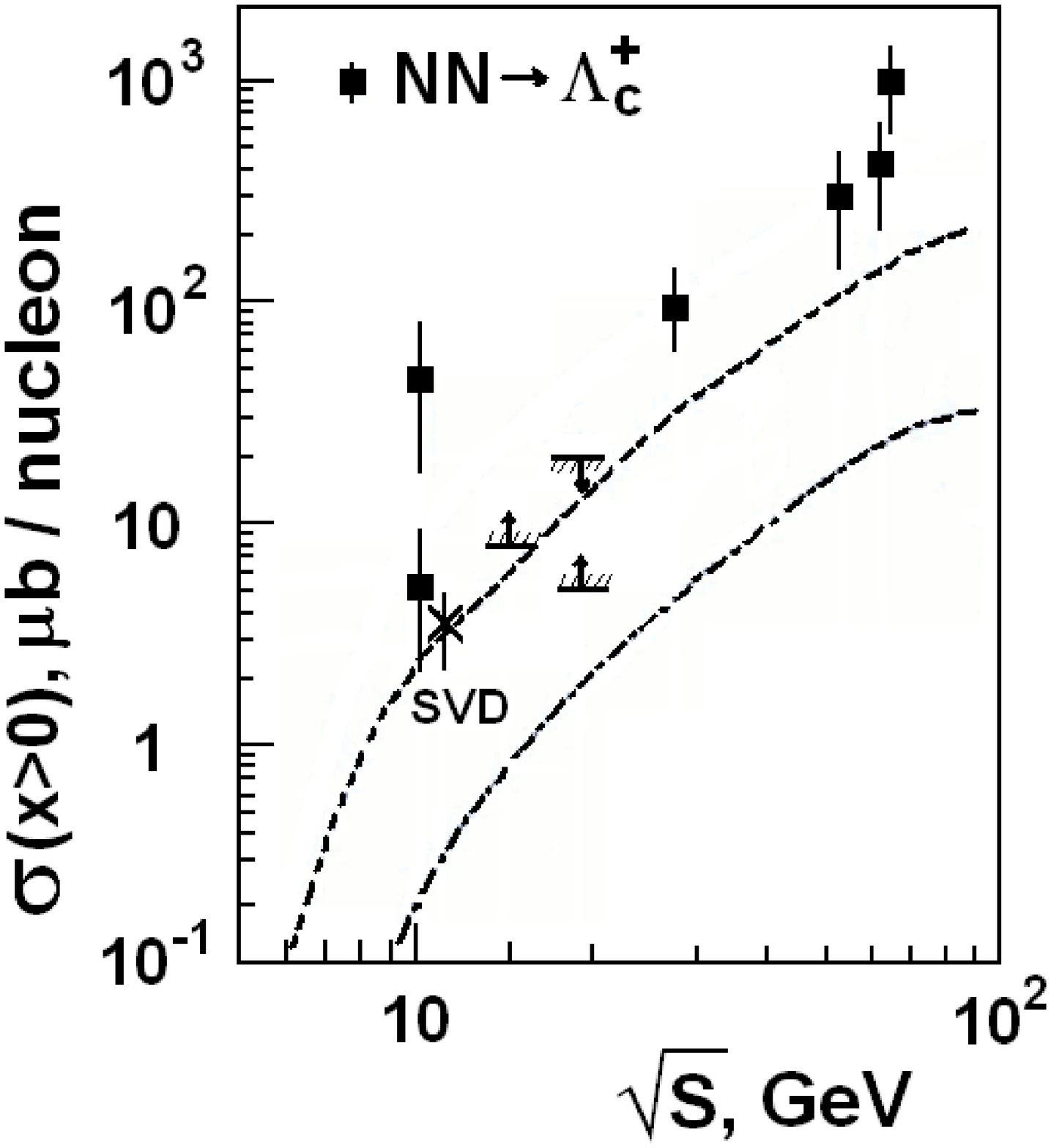}
\caption{ Left: The total cross section for charm production in pA-interactions
from \cite{Astratyan} - \cite{Aleev1}, \cite{Andronic1}. Extrapolation (solid line)
was carried out without points from IHEP:
$\bigcirc$ - SVD-2 experiment (see above),  $\bigtriangledown$ - beam-dump experiment
with muon absorber \cite{Astratyan}, $\sqcup$ - SCAT bubble chamber 
experiment \cite{Ammosov} and $\bigtriangleup$
-the experiment with BIS-2 spectrometer \cite{Aleev1}.  
Other lines are from various models \cite{Andronic1}.
Right: The inclusive cross section for $\Lambda ^+ _c$ baryon production
at x$_F$ > 0. The experimental data are from \cite{Aleev3}, \cite{Bailay} - \cite{Basile}
 (CERN, FNAL, BIS-2),
point (X) is from our experiment. Dashed lines are for predictions of two versions
of QCD model \cite{Aleev3}.   }
\label{fig-9}       
\end{figure*}
\item  The cross sections for $\Lambda ^+ _c$ production (Fig. 9 right) at the collision c.m. energy
$\surd s$ > 30 GeV contradict the total cross sections for the open charm production
(Fig. 9 left). The experimental cross sections for $\Lambda ^+ _c$ are extraordinarily large in this area.
\item The contributions of inclusive charmed particle cross-sections into the total
one vary at lower collision energies.
\end{itemize}

\section{Acknowledgements}

We appreciate IHEP (Protvino) staff for active and long-term joint work in the experiment
at U-70 accelerator dedicated to the charmed particles production at low energies.

%
%
%

\end{document}